# Pseudo spin-orbit coupling of Dirac particles in graphene spintronics

S. G. Tan[1], M. B. A. Jalil[2], Dax Enshan Koh[1], Hwee Kuan Lee[3], Y. H. Wu[2]

[1] Data Storage Institute, DSI Building, 5 Engineering Drive 1, (Off Kent Ridge Crescent, National University of Singapore) Singapore 117608

[2] Information Storage Materials Laboratory, ECE Dept., National University of Singapore, 4 Engineering Drive 3, Singapore 117576

[3] Bioinformatics Institute, 30 Biopolis Street, #07-01 Matrix, Singapore 138671

**Abstract**

We study the pseudo spin-orbital (SO) effects experienced by massive Dirac particles in graphene, which can potentially be of a larger magnitude compared to the conventional Rashba SO effects experienced by particles in a 2DEG semiconductor heterostructure. In order to generate a uniform vertical pseudo SO field, we propose an artificial atomic structure, consisting of a graphene ring and a charged nanodot at the center which produces a large radial electric field. In this structure, a large pseudo SO coupling strength can be achieved by accelerating the Dirac particles around the ring, due to the small energy gap in graphene and the large radial electric field emanating from the charged nanodot. We discuss the theoretical possibility of harnessing the pseudo SO effects in mesoscopic applications, e.g. pseudo spin relaxation and switching.

Contact: SG Tan

Email: Tan_Seng_Ghee@dsi.a-star.edu.sg

Tel: 65-68748410

The study of metal-based spintronics has led to the development of spin valves[1,2] in magnetic recording sensors that utilize the giant magnetoresistance[3] (GMR) effect, as well as the magnetoresistive random access memory[4] (MRAM) that is based on the tunneling magnetoresistance (TMR) effect. It has also led to the elucidation of the phenomenon of spin transfer excitations[5-7], and the demonstration of current-induced magnetization switching (CIMS) and spin torque oscillations in magnetic devices. On the other hand, in semiconductor-based spintronics, a key advantage is the presence of large Rashba and Dresselhaus spin-orbital[8-13] (SO) effects, which couple the spin and motion of particles (electrons or holes). They can thus be utilized to control electron spin in spintronic applications, e.g. in inducing spin precession of electron as it traverses across a two-dimensional electron gas (2DEG) conduction channel. Although experiments[14-16] have confirmed the working principles of such device propositions, major practical problems remain. These include low conductance modulation, stringent operational conditions such as low temperature and single-mode ballistic transport, and lack of controllability due to the non-uniformity of the Rashba and Dresselhaus constants in 2DEG heterostuctures. Additionally, it has been theoretically predicted that spin-orbit coupling in the presence of a longitudinal electric field in a 2DEG can generate a pure spin current in the direction transverse to the applied field, i.e. the spin Hall effect[17,18]. Recently, the theoretical prediction has been borne out by experimental observation of the intrinsic spin Hall effect induced by SO coupling[19]. Thus, a strong SO coupling will be a key requirement in proposed spin current sources based on the spin Hall effect. This may be achieved either by having a large electric field, e.g. at the interfaces of a 2DEG heterostructure, or by reducing the energy (mass) gap of the material. The latter suggests the use of monolayer or bilayer graphene[20-22] to enhance the analogous pseudo SO coupling effect, due to the small energy gap of Dirac particles here. In this article, we investigate the pseudo SO effects of particles in monolayer or bilayer graphene[21-23], in the relativistic and nonrelativistic limits. We show that a uniform pseudo SO coupling can be achieved in a graphene ring geometry, of a magnitude which cannot be attained by means of the conventional Rashba and Dresselhaus SO coupling in semiconductors. Finally, we propose how these pseudo spin-orbital effects can be utilized in device applications e.g. in pseudo spin switching.

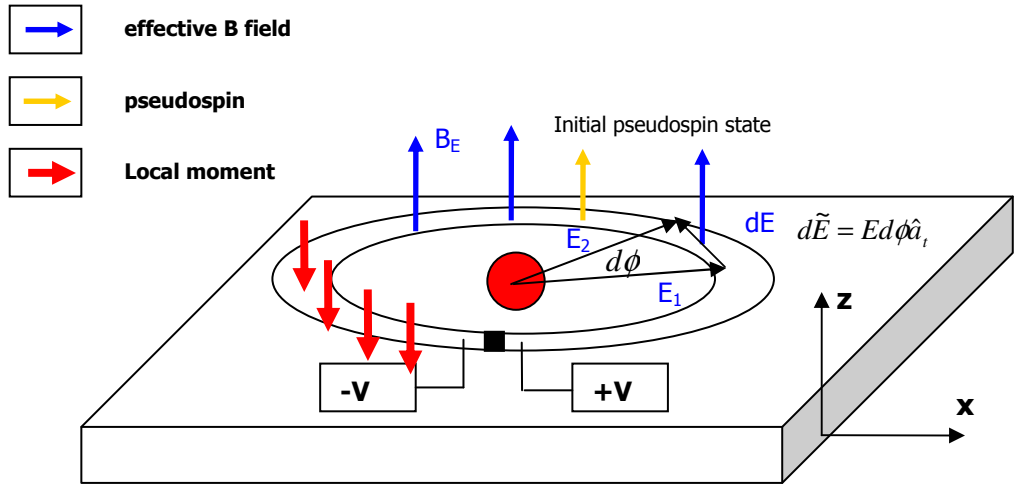

Figure 1. Graphene ring-and-dot nanostructure with large pseudo spin-orbit coupling suitable for applications in graphene spintronics

Figure 1 shows a graphene ring encircling a nanodot, both of which are of the order of several nanometers in size. Charges are trapped on the nanodot so as to produce a large radial electric field. The carriers in graphene, which act as Dirac particles near the K points, are accelerated around the ring to mimic the orbital motion of electrons around the nucleus of an atom. The sizable radial electric field, together with the small energy gap in graphene, translates into an extremely large pseudo SO coupling (where the pseudospin index indicates the sublattice contribution rather than real electron spin). The origin of the pseudo SO effect is analogous to the usual SO effect experienced by electrons moving in an electric field, i.e., a Lorentz transformation to the rest frame of the Dirac particle moving in a one-dimensional path around the circumference of the ring, produces a momentum dependent magnetic field which acts on the pseudospin of the particle. Mathematically, the SO term arises naturally from the Dirac equation in the non-relativistic limit (as will be shown below). Due to the geometry of the particle motion and the radial electric field, the effective magnetic field will be uniformly in the "vertical" direction (a vertical field is one which favors electron waves that are coupled between nearest sites belonging to the so-called A sublattice). The small energy gap in graphene (of the order of meV) means that the pseudo SO effects in the non-relativistic limit would be much larger compared to the Rashba or Dresselhaus SO effects whose effective energy gap in semiconductors is still much larger. In the presence of an electromagnetic field characterized by the four-potential $A_\mu = \left(\frac{\phi}{c}, -\mathbf{A}\right)$, where $\phi$ is the scalar potential and $\mathbf{A}$ is the vector potential, the Dirac[23,24] equation can be written in covariant form as:

$$(i\hbar\gamma^\mu D_\mu - mc)\psi = 0, \tag{1}$$

where $m$ is the rest mass of the electron, $\gamma^\mu$ are the Dirac matrices given by $\gamma^0 = \begin{pmatrix} I & 0 \\ 0 & -I \end{pmatrix}$ and $\gamma^j = \begin{pmatrix} 0 & \sigma^j \\ -\sigma^j & 0 \end{pmatrix}$, with $\sigma^j$ being the Pauli matrices, and $D_\mu = \left(\partial_\mu + \frac{ie}{\hbar}A_\mu\right)$ is the minimal coupling form of the four-momentum of the electron. Here, the Dirac matrices obey the commutation rule of Dirac algebra, namely $\{\gamma^\mu, \gamma^\nu\} = 2g^{\mu\nu}$, where $\text{diag}[g^{\mu\nu}] = (1, -1, -1, -1)$ is the Minkowski metric. Multiplying to the left of Eq. (1) by $i\hbar\gamma^\nu D_\nu + mc$, one obtains $\left(\hbar^2 g^{\nu\mu} D_\nu D_\mu + \frac{\hbar^2}{2}\gamma^\mu\gamma^\nu [D_\mu, D_\nu] + m^2c^2\right)\psi = 0$. From the definition of $D_\mu$, the commutator $[D_\mu, D_\nu]$ is equal to $\frac{ie}{\hbar}F_{\mu\nu}$, where $F_{\mu\nu} = \partial_\mu A_\nu - \partial_\nu A_\mu$ is the electromagnetic field tensor. Denoting $\Sigma^{\mu\nu} = \frac{i}{2}[\gamma^\mu, \gamma^\nu]$, one arrives at

$$\left(\hbar^2 D^\mu D_\mu + \frac{\hbar e}{2}\Sigma^{\mu\nu} F_{\mu\nu} + m^2c^2\right)\psi = 0. \tag{2}$$

In matrix representation, $F_{\mu\nu}$ and $\Sigma^{\mu\nu}$ are given by

$$[F_{\mu\nu}] = \begin{pmatrix} 0 & E^1/c & E^2/c & E^3/c \\ -E^1/c & 0 & -B^3 & B^2 \\ -E^2/c & B^3 & 0 & -B^1 \\ -E^3/c & -B^2 & B^1 & 0 \end{pmatrix} \text{ and } [\Sigma^{\mu\nu}] = \begin{pmatrix} 0 & i\alpha^1 & i\alpha^2 & i\alpha^3 \\ -i\alpha^1 & 0 & \sigma^3 & -\sigma^2 \\ -i\alpha^2 & -\sigma^3 & 0 & \sigma^1 \\ -i\alpha^3 & \sigma^2 & -\sigma^1 & 0 \end{pmatrix}, \text{ where}$$

$E^i$ and $B^i$ are components of the electric field $\mathbf{E}$ and the magnetic field $\mathbf{B}$, and $\alpha^j = \gamma^0 \gamma^j$. In the absence of an external magnetic field, i.e. when $\mathbf{B} = 0$, Eq. (2) can be expanded to give

$$\left[ (E - e\phi)^2 - c^2 p^2 - ie\hbar c \boldsymbol{\alpha} \cdot \mathbf{E} - m^2 c^4 \right] \psi = 0, \tag{3}$$

where $E = i\hbar c \partial_0$ is the energy of the electron, $\boldsymbol{\alpha} = (\alpha^1, \alpha^2, \alpha^3)$, and $p_j = \dfrac{\hbar}{i} \partial_j$ are the components of the momentum $\mathbf{p}$ of the electron. If we write the vector $\psi$ as a two-dimensional column vector, i.e $\psi = (\chi, \vartheta)^T$, where $\chi$ and $\vartheta$ are both two-dimensional vectors, Eq. (1) can then be expressed as

$$\begin{pmatrix} (E/c - e\phi/c) - mc & -\boldsymbol{\sigma} \cdot \mathbf{p} \\ \boldsymbol{\sigma} \cdot \mathbf{p} & -(E/c - e\phi/c) - mc \end{pmatrix} \begin{pmatrix} \chi \\ \vartheta \end{pmatrix} = 0, \tag{4}$$

whilst Eq. (3) becomes

$$\begin{pmatrix} (E - e\phi)^2 - c^2 p^2 - m^2 c^4 & -ie\hbar c \boldsymbol{\sigma} \cdot \mathbf{E} \\ -ie\hbar c \boldsymbol{\sigma} \cdot \mathbf{E} & (E - e\phi)^2 - c^2 p^2 - m^2 c^4 \end{pmatrix} \begin{pmatrix} \chi \\ \vartheta \end{pmatrix} = 0. \tag{5}$$

In the above, $\boldsymbol{\sigma} = (\sigma^1, \sigma^2, \sigma^3)$ is the vector of Pauli matrices. From Eq. (4), we obtain

$\vartheta = \dfrac{\boldsymbol{\sigma} \cdot \mathbf{p}}{E/c - e\phi/c + mc} \chi$. This is then substituted into the relation

$\left[ (E - e\phi)^2 - c^2 p^2 - m^2 c^4 \right] \chi - ie\hbar c (\boldsymbol{\sigma} \cdot \mathbf{E}) \vartheta$ which follows from Eq. (5), and using the identity

$(\boldsymbol{\sigma} \cdot \mathbf{E})(\boldsymbol{\sigma} \cdot \mathbf{p}) = i\boldsymbol{\sigma} \cdot (\mathbf{E} \times \mathbf{p}) + \mathbf{E} \cdot \mathbf{p}$, one then obtains:

$$\left[ (E - e\phi)^2 - c^2 p^2 - m^2 c^4 - ie\hbar c^2 \frac{\mathbf{E} \cdot \mathbf{p}}{E - e\phi + mc^2} + e\hbar c^2 \frac{\boldsymbol{\sigma} \cdot (\mathbf{E} \times \mathbf{p})}{E - e\phi + mc^2} \right] \chi = 0 \tag{6}$$

In the non-relativistic limit, i.e. assuming a small $\varepsilon = E - mc^2$ and $e\phi$, Eq. (6) reduces to the following Hamiltonian:

$$\left[\frac{p^2}{2m} + e\phi + \frac{\hbar e}{4m^2c^2}\boldsymbol{\sigma}\cdot(\mathbf{p}\times\mathbf{E}) + \frac{i\hbar e}{4m^2c^2}\mathbf{E}\cdot\mathbf{p}\right]\chi = \varepsilon\chi, \quad (7)$$

where the third term denotes the SO coupling. The SO term can be expressed as

$$H_{SO} = \frac{\hbar e}{4m^2c^2}\boldsymbol{\sigma}\cdot(\mathbf{p}\times\mathbf{E}) = Ge\boldsymbol{\sigma}\cdot\left[\frac{\partial \mathbf{r}}{\partial t}\times\mathbf{E}\right] = Ge\boldsymbol{\sigma}\cdot\left[\frac{d}{dt}(\mathbf{r}\times\mathbf{E}) - \mathbf{r}\times\frac{\partial \mathbf{E}}{\partial t}\right], \quad (8)$$

where $G = \frac{\hbar}{4mc^2}$ is a time constant (for electrons, $G = 3.2\times10^{-22}$ s) which denotes the SO strength and is inversely proportional to the energy gap $\Delta = 2mc^2$ between a particle and a hole. For the ring structure of Fig. 1, the magnitude of the electric field is $|\mathbf{E}| = \frac{Ne}{4r^2\pi\varepsilon_0}$, where $r$ is the ring radius, $N$ is the number of discrete charges in the nanodot and $\varepsilon_0$ = 8.85x10$^{-12}$ C$^2$J$^{-1}$m$^{-1}$ is the permittivity of free space. Since the electron's displacement and the E-field are both pointing radially outwards from the centre of the ring, the $(\mathbf{r}\times\mathbf{E})$ term in Eq. (8) vanishes. The other term can be expressed as $\mathbf{r}\times\frac{\partial \mathbf{E}}{\partial t} = \left(r\hat{\mathbf{a}}_r \times |\mathbf{E}|\frac{v}{r}\hat{\mathbf{a}}_t\right)$, where $\hat{\mathbf{a}}_r$ and $\hat{\mathbf{a}}_t$ are the radial and tangent unit vectors respectively, and $v$ is the velocity of the electron. Thus, Eq. (8) reduces to:

$$H_{SO} = -Ge|\mathbf{E}|v\boldsymbol{\sigma}\cdot(\hat{\mathbf{a}}_r\times\hat{\mathbf{a}}_t). \quad (9)$$

From Eq. (9), we find that particles circulating the ring in the clockwise (counter-clockwise) direction will experience an effective magnetic field that points uniformly upward (downward) throughout the ring. Assuming a ring radius of $r$ = 10 nm, and a moderately high particle tangential velocity of $v$ = 10$^4$ ms$^{-1}$. At the ring radius, we have an E-field strength of 1.4 x N x 10$^7$ Vm$^{-1}$, which corresponds to a SO coupling energy of 4.5 x N x 10$^{-8}$ meV, if the ring material has a vacuum energy gap of ($\Delta$ = 2m$_e$c$^2$=1 MeV). Thus, comparing with the conventional Rashba SO coupling in a 2DEG made of III-V semiconductor materials, where the SO coupling energy typically ranges from 1-4 meV, we require an impractically large charge accumulation of 10$^8$ electrons on the nanodot to produce an equivalent SO coupling energy. The situation would be vastly different for pseudo-SO effect in rings made of graphene-like materials. Experiments have shown that the energy gap experienced by Dirac particles in monolayer graphene grown on SiC[25-26] is approximately 0.26 eV, and can be as low as 0.04 eV, with proper engineering. This corresponds to a pseudo-SOC energy of 0.17x*N* to 1.1x*N* meV. Recent theoretical computations have indicated that $\Delta$ can be as low as 10 meV in bilayer and trilayer graphene[27]. Thus, even with one discrete charge on the nanodot, the pseudo-SO strength matches the Rashba effect in 2DEG structures. In addition, the energy gaps of graphene have also been shown to be tunable, which implies that the SO coupling strength can be tuned not only by charging up the nanodot, but also modifying the energy gap[28] in graphene. Besides, there are other device parameters that can be adjusted to maximize the SOC strength. For instance, advances in nanofabrication can allow the ring radius to be further reduced, thus increasing the pseudo SO energy significantly because of the $(1/r^2)$-dependence of the *E*-field. Particle velocity around the graphene ring can also be stepped up to increase the pseudo SO energy via $\partial E/\partial t$.

We will now investigate the effects of pseudo SO coupling on pseudo spin relaxation. As can be seen from Eq. (8), a Dirac particle in the graphene ring experiences an effective pseudo magnetic field $\mathbf{B}_E = (\mathbf{p} \times \mathbf{E})$, which couples directly to its momentum vector, thus preserving time-reversal symmetry. Such a field can be utilized for a controlled precession of pseudo-spin when coupled with appropriate momentum constraints (e.g. single mode one-dimensional ballistic transport), similar to gate bias-controlled spin precession via Rashba or Dresselhaus SO coupling in the Datta-Das spin transistor[11]. It has also been shown that the degree of electron precession about the effective SO magnetic fields is related to the spin relaxation to the effective field direction[29]. A similar phenomenon would occur in the case of pseudo spin in graphene under the influence of the pseudo SO effect. Dirac particles with energy just exceeding the small energy gap can be taken as non-relativistic. In the non-relativistic limit, the expectation energy as obtained from Eq. (7) is given by: $\left|\left\langle\chi\left|H_{SOC}\right|\chi\right\rangle\right|=\hbar epE\sin\eta\cos\kappa/4m^2c^2\equiv\hbar\omega$, where $\omega$ is the pseudo spin precession frequency, and $\kappa$ and $\eta$ are the angles between $\boldsymbol{\sigma}$ and $(\mathbf{p}\times\mathbf{E})$, and $\mathbf{p}$ and $\mathbf{E}$, respectively. For simplicity, we consider the case where $\sin\eta=\cos\kappa=1$. The average particle velocity can be approximated as $\left\langle\chi|v|\chi\right\rangle=p/m$, neglecting the acceleration due to the E field, and higher-order terms e.g. due to Zitterbewegung motion. In a semiclassical manner, the precession angle $\Omega$ per unit travelling distance is given by:

$$\Omega=\frac{\omega}{\left\langle\chi\left|v_\mu\right|\chi\right\rangle}=\frac{eE}{4mc^2}, \quad (10)$$

which is independent of its velocity. However, due to the small energy gap $mc^2$ in graphene and the tunable pseudo SO energy in the graphene ring, $\Omega$ for pseudo spin in graphene can potentially be much larger than that for electron spins in the semiconductor quantum well. Since $\Omega$ is related to the spin relaxation efficiency, it therefore means that pseudo spins will more readily undergo adiabatic relaxation to the local pseudo SO field, compared to physical spins subject to the Rashba effect. Hence, for instance, the intrinsic spin Hall effect, which is postulated to arise from adiabatic relaxation of spins to the local SO field direction, will be much more prominent for the case of pseudo spins in graphene. It would be instructive to analyze the pseudo spin relaxation in the relativistic limit, as this phenomenon is achievable in graphene via narrowing the energy gap. Without taking the non-relativistic limit as we did in Eq. (7), Eq. (6) may be written as

$$\left[E^2-m^2c^4-p^2c^2-\frac{ie\hbar c}{p_0+mc}\mathbf{E}\cdot\mathbf{p}-\frac{e\hbar}{p_0+mc}\boldsymbol{\sigma}\cdot(\mathbf{E}\times\mathbf{p})c\right]\chi=0, \quad (11)$$

where $p_0=(E/c)$. We have neglected the scalar potential $\phi$, which is assumed to be small compared to the energy $E$, which is large in the relativistic limit. Taking the expectation value of the spin-orbit related operator, $\left\langle\chi\left|H_{SOC}\right|\chi\right\rangle$ yields $E_R=\dfrac{\hbar ep_\mu E\sin\eta\cos\kappa}{p_0+mc}c$. Assuming as before that $\sin\eta = \cos\kappa = 1$, we obtain the precession frequency to be $\hbar\omega = \sqrt{\dfrac{\hbar epEc}{p_0 + mc}}$.

Once again, assuming the simplest form of one-dimensional particle motion, i.e. neglecting the effects of the E-field and Zitterbewegung, the average velocity is given $\langle\chi|v_\mu|\chi\rangle = \dfrac{cp}{\sqrt{p^2 + m^2c^2}}$.

Thus, the precession angle for a unit of particle travel length in the relativistic regime is given by:

$$\Omega = \sqrt{\dfrac{eEp_0^2}{\hbar(p_0 + mc)pc}} \ . \qquad (12)$$

For simplicity, we consider the case where the kinetic energy $pc$ is relatively small compared to the energy gap $mc^2$, so that the above reduces to $\Omega = \sqrt{\dfrac{eEm}{2\hbar p}}$. Thus, in this instance, increasing the momentum would reduce the precessional angle, thereby suppressing pseudo spin relaxation, in contrast with the non-relativistic case in Eq. (10), where $\Omega$ is independent of momentum. These results indicate that pseudo spin relaxation can be effectively suppressed in the relativistic limit by increasing the particle momentum.

Considering the fact that pseudospin magnetism might exist in graphene[30], we envisage the use of our ring structure as a memory or logic device whose binary states are based on the pseudo spin moment. We can then utilize the pseudo SO field to achieve current-induced switching of this memory. For example, suppose that the graphene ring has an initial pseudospin moment in the negative vertical direction (as shown in Fig. 1). To switch the pseudospin moment, we pass the Dirac particles in the anticlockwise direction around the ring by applying a voltage of the right polarity. This would produce an effective pseudo SO field $\mathbf{B_E}$ in the positive vertical direction, which would align the pseudo spin of the particles oppostite to the pseudo spin moment of the ring, and in turn causing the latter to switch when the current is sufficiently large. The current induced switching of pseudo spin magnetism potentially has advantages over the current-induced magnetization switching (CIMS) in metallic ferromagnetic multilayers. For instance, the required current density for switching the pseudo spin magnetization in either direction would be symmetrical in magnitude. By contrast, conventional CIMS involves the passage of carriers through a reference (or "pinned") magnetic layer to switch the other ("free") magnetic layer to the parallel direction, and relies on the reflection of carriers to induce switching in the antiparallel direction. Due to the different mechanisms being involved, conventional CIMS suffers from asymmetrical switching current for free layer moments of different polarity. Furthermore, the proposed graphene ring memory involves direct injection of carriers, without the need for them to be polarized by passage across an additional pinned layer. Thus, the CIMS analogue in graphene can sidestep practical issues afflicting the conventional CIMS in magnetic multilayers and magnetic tunnel junctions (MTJ), e.g. spin decoherence due to transport and spin flip in the different layers and interfaces, which seriously degrades the efficiency of spin transfer.

In summary, we have studied the pseudo SO coupling effect in a proposed graphene ring-and-dot nanostructure. Charge accumulation in the nanodot produces a radial electric field, while particle circulation around the ring mimics the orbital motion of electrons about the atomic nucleus. We have analyzed the pseudo spin precessional frequency, and hence the pseudo spin relaxation in both the relativistic and non-relativistic cases. Finally, the graphene ring that we have proposed can be used as a memory/logic device based on pseudo spin magnetism and pseudo spin CIMS in the graphene ring, which has practical advantages in terms of symmetric switching field, and the absence of pinned layer. From the technological point of view, the proposed device can readily achieve downward size scalability and increased density, since the pseudo SO effect is enhanced in smaller rings. The fact that graphene provides a natural bench-top setting for massive Dirac particle-like behavior in a condensed matter system paves the way for utilization of relativistic effects for applications. Our studies have analyzed these effects in the context of pseudo spintronics, i.e. pseudo spin-orbital effects and pseudo spin switching and relaxation.

**ACKNOWLEDGEMENT**

We thank the National University of Singapore (NUS) and the Agency for Science, Technology, and Research of Singapore for funding fundamental research in spintronics and graphene physics.